\documentclass[twocolumn,prl,showpacs,preprintnumbers,amsmath,amssymb]{revtex4}

\usepackage{graphicx}
\usepackage{dcolumn}
\usepackage{bm}

\begin{document}

\title{
Zeeman levels with exotic field dependence\\ in the high field phase of a 
$\bm S=1$ Heisenberg antiferromagnetic chain
}

\author{M. Hagiwara$^1$, Z. Honda$^2$, K. Katsumata$^3$, A. K. 
Kolezhuk$^{4,5}$, and H.-J. Mikeska$^4$}
\affiliation{
$^1$RIKEN(The Institute of Physical and Chemical Research), Wako, Saitama 
351-0198, Japan \\ $^2$Faculty of Engineering, Saitama University, Saitama 338-8570, 
Japan \\ $^3$RIKEN Harima Institute, Mikazuki, Sayo, Hyogo 679-5148, Japan \\ 
$^4$Institut f\"ur Theoretische Physik, Universit\"at Hannover, 
Appelstra{\ss}e 2, 30167, Hannover, Germany \\ $^5$Institute of 
Magnetism, 36(B) Vernadskii avenue, 03142 Kiev, Ukraine}

\date{\today}

\begin{abstract}
We have performed electron spin resonance 
 measurements over a wide frequency and magnetic field range on
a single crystal of the $S=1$ quasi-one-dimensional Heisenberg
antiferromagnet $\rm Ni(C_5H_{14}N_2)_2N_3(PF_6)$.  We observed gapped 
excitation branches above the critical field ($H_{\rm c}$) where the 
Haldane gap closes. 
These branches are analyzed by a phenomenological field-theory
using complex-field $\phi^{4}$ model. A satisfactory agreement between
experiment and theory is obtained.  
\end{abstract}

\maketitle


Recently, field-induced phenomena in quantum spin systems have 
attracted 
considerable interest.  These include
magnetization plateaus~\cite{narumi,shiramura,kageyama} and field-induced 
long-range order (LRO)~\cite{honda1,honda2,chen}.   
Gapped one-dimensional (1D) spin systems  subject to an  external 
magnetic field strong enough to close the gap are driven into a 
new phase. 
When the system has XY or Heisenberg symmetry,  this phase is critical and its
low-energy physics is described by the Tomonaga-Luttinger liquid~\cite{chitra},
whose  elementary excitations are of the particle-hole type (spinon
pairs). 
The high-energy physics which cannot be described 
using this picture has been 
recently investigated  
experimentally \cite{orendac,zheludev,ruegg} as well as
theoretically \cite{KolezhukMikeska}.

\begin{figure}
\includegraphics[width=6cm,clip]{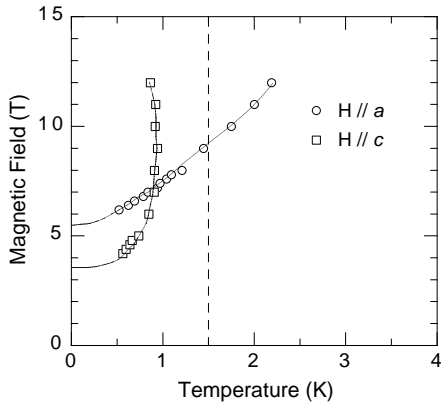}
\caption{The magnetic field versus temperature phase diagram of NDMAP;  
solid lines are guides to the eye (figure from \protect\cite{honda1}).
The ESR measurements were performed along the dashed line. } 
\label{fig:htphase}
\end{figure}

One-dimensional Heisenberg antiferromagnets (HAFs) with integer spin
are typically gapped 
and have a disordered
spin liquid singlet ground state~\cite{haldane}. The first excited
state is a triplet and one of the 
Zeeman-split triplet branches goes down in magnetic field. At some
critical field $H_{\rm c}$ the gap closes, a qualitatively new ground
state emerges and magnetism recovers.  Above $H_{\rm c}$, 
a LRO is expected to occur in quasi-1D integer-spin HAFs owing to interchain
interaction.
Such a field-induced LRO in the $S=1$
quasi-1D HAF was found recently by specific heat measurements on a
single crystal of $\rm Ni(C_5 H_{14} N_2)_2 N_3(PF_6)$, abbreviated
NDMAP~\cite{honda1,honda2}.  Figure~\ref{fig:htphase} shows the
magnetic field  versus temperature phase diagram of NDMAP.
The regions below and above the boundary denoted as $H_{\rm LRO}(T)$
correspond to the Haldane disordered phase, and to the LRO phase,
respectively.  
The critical field
$H_{\rm c}$ in the present paper is defined as $H_{\rm LRO}$ extrapolated
to $T=0$.

Quite recently, 
three distinct excitation branches were observed in inelastic neutron
scattering (INS) experiments on NDMAP 
in the
LRO phase above $H_{\rm c}$ for $H \parallel a$, 
and a satisfactory desription within a phenomenological field
theory was obtained \cite{zheludev}. 
This observation is remarkable because
one would expect two  branches from the conventional spin-wave theory.

Electron spin resonance (ESR) is one of the powerful methods to investigate
magnetic excitations, especially at low energies with a much better
energy resolution than INS.  In Ref.\ \onlinecite{honda3}, we 
observed ESR signals in a single crystal of NDMAP and interpreted
the results as the coexistence of 
1D and 3D excitations, because no satisfactory description was
available above $H_{\rm c}$ at that time.

In the present paper we extend the ESR study of NDMAP to a much 
wider frequency range from which we obtain new results.
 A nontrivial field dependence of the resonance
modes, very different from that expected from a conventional
spin-wave theory, is observed.  
We compare the experimental results with model
calculations based on the field theory used in the analysis of the
recent INS experiments \cite{zheludev}, and obtain a good agreement
with the theory.  
 

First we summarize the crystal and magnetic properties of NDMAP.  This
compound crystallizes in the orthorhombic system (space group
$Pnmn$)~\cite{monfort}.
The $\rm
Ni^{2+}$ ions are bridged by azido ions  
forming 
antiferromagnetic chains along the $c$ axis.
All the $\rm Ni$
sites in a chain are crystallographically equivalent, so that
no staggered components of the magnetic moments are retained in the
ground state below $H_{\rm c}$.  
The physics in NDMAP is, therefore, essentially different 
from that in another typical Haldane chain material 
$\rm Ni(C_2H_8N_2)_2NO_2(ClO_4)$, abbreviated NENP, extensively
studied by ESR \cite{lu-sieling-brill}. These studies showed 
that the Haldane gap did not close at H$_c$ due to a staggering of the 
local crystal fields for
Ni$^{2+}$\cite{MitraHalperin94,SakaiShiba94}. Consequently, in NENP no 
clear transition from the disordered to LRO phase is observed and the 
ESR transitions from the ground state are always allowed.

From the analysis of the magnetic susceptibility data, the following 
values are obtained~\cite{honda1}: $J/k_{\rm B}=30.0$~K, $D/J=0.3$, 
$g_{\parallel}=2.10$ and $g_{\perp}=2.17$, where $J$ is the intrachain 
exchange interaction constant and $g_{\parallel}$ and 
$g_{\perp}$ are the $g$ values parallel and perpendicular to the chain, 
respectively.  From the INS measurements done 
on single crystals of deuterated NDMAP~\cite{zheludev2}  the 
following parameters were determined: 
$J=2.85$~meV($=33.1$~K), $J^{\prime}_x=3.5\times10^{-4}$~meV
($=4.1\times10^{-3}$~K), 
$J^{\prime}_y=1.8\times10^{-3}$~meV ($=2.1\times10^{-2}$~K) and
$D=0.70$~meV ($=8.1$~K), 
where 
$J^{\prime}_x$ and $J^{\prime}_y$ are the interchain exchange 
interactions along the $a$ and $b$ axes, respectively.


The single crystals of NDMAP were grown in the same manner as that 
described in Ref.\onlinecite{honda1}. The ESR
measurements were performed using a millimeter vector network analyzer
with extensions up to about 700~GHz from the AB Millimeter, France
and a superconducting magnet up to 16~T from the Oxford Instruments,
UK.  We performed the ESR measurements in $H\parallel a$ and
$H\parallel c$ geometries at the lowest temperature available with our
spectrometer ($1.5$~K).   Faraday and Voigt configurations were used 
for $H\parallel a$ and $H\parallel c$, respectively. As shown 
in Fig.~\ref{fig:htphase}, the external magnetic field of $\approx 9$~T for
$H\parallel a$ brings the system into the LRO phase at $1.5$~K.


\begin{figure}
\includegraphics[width=6cm,clip]{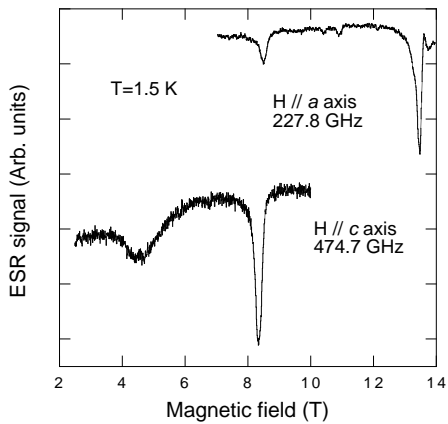}
\caption{An example of ESR spectra for $H\parallel a$ and $H\parallel
c$ obtained at $1.5$~K. } 
\label{fig:spectra}
\end{figure}

\begin{figure}
\includegraphics[width=6cm,clip]{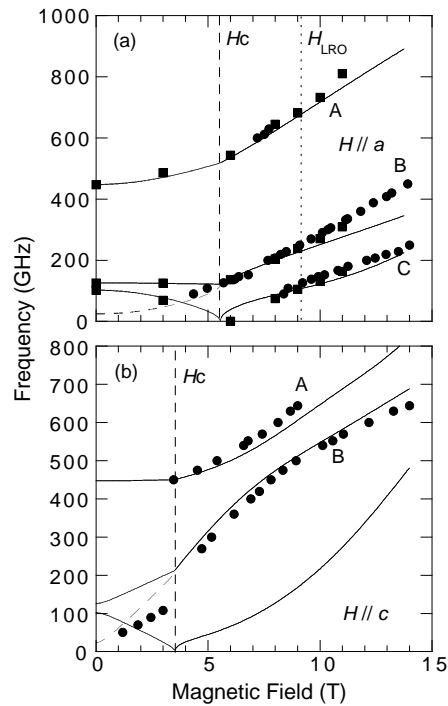}
\caption{The ESR resonance fields are plotted with solid circles in
the frequency versus magnetic field plane for (a) $H\parallel a$ 
(the field perpendicular to the chain) and (b) $H\parallel c$  (the field
parallel to the
chain), respectively.  The broken vertical line is the critical field
$H_{\rm c}$ at $T=0$~K.  The solid lines are fits based on the
field-theoretical approach.  The thin broken line below $H_{\rm c}$
denotes transition between the lowest two excited states.  In the
upper panel, the solid squares are taken from the neutron inelastic
measurements\cite{zheludev} at the wave vector $q=\pi$.  The dotted
vertical line denotes the boundary between the disordered and the LRO
phases at $T=1.5$~K (see \ Fig.\ \ref{fig:htphase}).  }
\label{fig:f-H}
\end{figure}

Figure~\ref{fig:spectra} shows a typical example of ESR spectra observed for 
$H\parallel a$ and $H\parallel c$.   The signal at about $13.5$~T for
$H\parallel a$ comes from the LRO phase. 
In Fig.~\ref{fig:f-H}(a), the resonance frequencies for $H\parallel a$
as a function of magnetic field are shown with solid circles.  We also
plot with solid squares the gap energies obtained from the recent INS
measurements at 30~mK~\cite{zheludev}.  In the figure, the vertical
broken line and the vertical dotted line denote the $T=0$ critical
field $H_{c}$ and the boundary field $H_{\rm LRO}$ between the
disordered and LRO phases at $1.5$~K, respectively.  It is obvious
that the slope of the branches changes at $H_{\rm c}$ and not at
$H_{\rm LRO}$.  It should be remarked that a similar change of slope
was observed in the quasi-1D material NENC \cite{orendac} and in the
3D weakly coupled dimer system $\rm TlCuCl_{3}$
\cite{ruegg,Matsumoto+02}.  The observed neutron data below $H_{\rm
c}$ represent the magnon triplet split by the single ion anisotropy in
the Haldane phase.  Below $H_{\rm c}$, only temperature-induced
ESR transitions between the magnon branches are allowed.  Only one ESR
line, corresponding to the transition between the lowest two branches,
was observed near the critical field below $H_{\rm c}$.  Above $H_{\rm
c}$, three ESR lines were observed. Their positions coincide with the
INS data for the gaps at $q=\pi$, confirming that they correspond to
the transitions from the ground state. It is worthwhile to note that
these lines appear at $H>H_{\rm c}$ and not at $H>H_{\rm LRO}$ as one
would naively expect, and that there is no anomaly in the behavior of
resonance lines at $H_{\rm LRO}$. 
Indeed, the gap of the lowest magnon
branch at $q=\pi$ closes at $H=H_{\rm c}$ at any temperature, causing
``condensation'' of $q=\pi$ magnons into the ground state at $H>H_{\rm
c}$.
The amplitude of the $q=\pi$ component becomes nonzero
immediately above $H_{\rm c}$, making the ESR transitions from the
ground state to the excited ones at $q=\pi$ possible.
The transversal staggered LRO, however, emerges only at the
higher field $H_{\rm LRO}$, because the phase
coherence of the condensate wave function,  directly related
to the LRO, is destroyed by thermal fluctuations in
the field range $H_{\rm c}<H<H_{\rm LRO}$.
When the phase coherence sets in at $H=H_{\rm LRO}$, there is no
peculiarity in the amplitude and thus no essential change in the ESR
signal.

Figure~\ref{fig:f-H}(b) shows a frequency versus field plot
for $H\parallel c$.  As in the $H\parallel a$ case, we observed only
one resonance line below $H_{\rm c}$ which corresponds to the
temperature-induced transitions between the lowest two magnon
branches.  Above $H_{\rm c}$, two ESR lines were observed.  For this
geometry, there is no LRO at $1.5$~K, so that the measurements were
conducted completely outside the LRO phase.  One of the remarkable features
in this figure is a change in slope at about 7~T, far above the
critical field $H_{\rm c}\simeq 3.5$~T, resulting from an avoided
crossing between the two branches (see below).


Now, we compare the experimental results with theoretical calculations
using a field-theoretical model \cite{zheludev} similar to that
proposed for dimerized $S=\frac{1}{2}$ chains in Ref.\
\onlinecite{K96}.  It is a Ginzburg-Landau-type theory written
in terms of a \emph{complex} triplet field $\bm{\phi}=\bm{A}+i\bm{B}$
which
is assumed to be small, $|\bm{\phi}|\ll 1$. 
For a dimerized $S=\frac{1}{2}$ system, the field $\bm{\phi}$
describes the dimer wave function
$|\bm{\phi}\rangle=(1-|\bm{\phi}|^{2})^{1/2}|s\rangle +\sum_{a=x,y,z} 
\phi_{a}|t_{a}\rangle$,
where $|s\rangle$ and $|t_{a}\rangle$ are  the singlet and
three triplet states of the dimer; it is a continuum analog of
the bond boson description used in \cite{Matsumoto+02}. 
The uniform and staggered magnetization are given by
 $\bm{M}=(\bm{\phi}^{*} \times \bm{\phi})$ and
$\bm{L}=(\bm{\phi}^{*}+\bm{\phi})(1-|\bm{\phi}|^{2})$, respectively.
The effective Lagrangian density in the continuum limit has the
following form:
\begin{eqnarray} 
\label{Leff-anis} 
{\mathcal L}&=& \hbar
(\bm{A}\cdot\partial_{t}\bm{B}-\bm{B}\cdot\partial_{t}\bm{A}) 
-(1/2)v^{2}(\partial_{x}\bm{A})^{2} \nonumber\\
&-&\sum_{\alpha}\{ 
m_{\alpha}A_{\alpha}^{2}
+\widetilde{m}_{\alpha}B_{\alpha}^{2} \}
+ 2\bm{H}\cdot(\bm{A}\times\bm{B})\\
&-&\lambda(\bm{A}^{2})^{2}-\lambda_{1}(\bm{A}^{2}\bm{B}^{2}) 
-\lambda_{2}(\bm{A}\cdot\bm{B})^{2}.\nonumber
\end{eqnarray}
This model, though derived for dimerized $S=\frac{1}{2}$ chains, may
also be used for other gapped 1D systems which are in the same
universality class, particularly for $S=1$ Haldane chains.
In this case, the quantities $m_{\alpha}$, $\widetilde{m_{\alpha}}$, and
$\lambda_{i}$ should be treated as phenomenological parameters;
the direct microscopic justification of this model for $S=1$ chains is
the subject of ongoing work.  The
spatial derivatives of $\bm{B}$ are omitted in (\ref{Leff-anis})
because they appear only in terms which are of the fourth order in
$\bm{A}$ and $\bm{B}$.
By integrating out the ``slave''
$\bm{B}$-field one obtains an effective real-field $\varphi^4$-type
theory similar to that of Affleck \cite{Affleck90-91}; actually,
Affleck's theory may be viewed as a special case of
(\ref{Leff-anis}) with isotropic $\widetilde{m}_{\alpha}=\widetilde{m}$
and with simplified interaction term $\lambda_{1,2}=0$. Another
special case $m_{\alpha}=\widetilde{m_{\alpha}}$ leads to the spectra
which for $H<H_{c}$ exactly coincide with those obtained in the approach of Tsvelik
\cite{Tsvelik90} who has proposed an effective theory involving three
massive Majorana fields, and also with the perturbative formulas of
Refs.\ \onlinecite{perturbative}. If one additionally assumes the
simplified  interaction potential $\lambda_{1,2}=0$, the theory
becomes equivalent to that of Mitra and Halperin \cite{MitraHalperin94}
who postulated a bosonic Lagrangian to match Tsvelik's results for the
field dependence of the gaps below $H_{c}$.

The solid lines in Figs.\, \ref{fig:f-H}(a) and (b) represent 
the best fit of the calculated energy gaps at $q=\pi$ to the
experimental data.  This fit is obtained using the following set of
model parameters (all values in meV): 
$m_{a}=0.5$, $m_{b}=0.8$, $m_{c}=5.29$, $\widetilde{m_{a}}=0.353$,
$\widetilde{m_{b}}=0.342$, $\widetilde{m_{c}}=0.647$, $\lambda=1$,
$\lambda_{1}=-\lambda_{2}=0.13$.  We obviously see a change in slope of
each excited branch and re-opening the energy gap of the lowest branch
at $H_{\rm c}$.  Below $H_{\rm c}\simeq 5.5$~T in
Fig.~\ref{fig:f-H}(a), three excitation branches are in a good
agreement with the neutron data.  The ESR transitions from the
ground to the excited states at $q=\pi$ are forbidden
because no staggered component exists in the ground state of this
compound.  In both figures, the 
ESR resonance points only appear near
$H_{\rm c}$, because these correspond to the transition between the
excited states and the intensity is related to the population of the
excited states at $1.5$~K. The thin broken line in each figure indicates
the transition between the lowest two magnon branches at $q=\pi$. For those
temperature-induced transitions, the agreement
between experiment and calculation is satisfactory for
$H\parallel a$  and is not good for $H \parallel c$.
Above $H_{\rm c}$, the calculated branches are close to the
experimental data near $H_{\rm c}$, but deviate from them at high
fields.  This is probably because the theory is applicable only for
the state with a small staggered order and thus only for 
fields which are not too far above $H_{\rm c}$.

In these calculations, we have taken into account the tilting of
the crystal field axis of the ${\rm Ni^{2+}}$ ions from the crystallographic
$c$ axis. The tilting angle is $15.9^{\circ}$.  In
Fig.~\ref{fig:f-H}(b), a change in slope at about 7~T observed in the
A and B branches are reproduced in the calculation.  Because of the tilting 
of the crystal field axis, the external magnetic field 
applied parallel to the $c$ axis mixes the wave functions of A
and B branches.  The origin of this change in slope is different from
that of the changes visible near the critical field.    
The observed ESR line
below $H_{\rm c}$ in Fig.~\ref{fig:f-H}(b) deviates from the
expected thin broken line, although a satisfactory agreement is
obtained for $H\parallel a$; the reason for this discrepancy is not
clear at present.   For $H\parallel c$,
we did not observe ESR signals corresponding to the lowest calculated mode
above $H_{\rm c}$; we believe this is caused by 
a strong damping of the
lowest magnon mode due to the interaction with domain walls, because 
our measurement was done at a high temperature. A
similar problem arose in INS measurements
\cite{Zheludev+02-PRL} in the $H\parallel a$ geometry done at $2$~K
where the lowest magnon mode appeared to be quasielastic, and 
only after lowering the temperature to $30$~mK the mode was
successfully observed \cite{zheludev}. The damping should be much
stronger in the $H\parallel c$ geometry, due to the lower 
energy and respectively higher density of domain walls.

Remarkably, the observed field dependence of the ESR modes for both
geometries is very different from that expected from a conventional
spin-wave theory.  Indeed, in a conventional (classical)
antiferromagnet with easy-plane anisotropy $D$ and external magnetic
field $H$ applied in the easy plane, there are two resonance modes
with the energies $\varepsilon_{1}=g\mu_{B}H$ and
$\varepsilon_{2}=2S\sqrt{2DJ}$.  For $H$ perpendicular to the easy
plane, one would classically expect to see one field-dependent
resonance mode with the energy
$\varepsilon_{2}'=\sqrt{(g\mu_{B}H)^2+\varepsilon_{2}^{2}}$, the
energy of the other mode being zero.  We have observed 
and
described theoretically
completely
different behaviors above $H_{\rm c}$ for both directions: in the
$H\parallel a$ case (in-plane geometry) the energies of all modes
increase approximately as $g\mu_{B}H$, and for $H\parallel c$ we
observe the (avoided) crossing of two modes whose energies behave
roughly as $g\mu_{B}H$ and $2g\mu_{B}H$.


In conclusion, we have performed ESR measurements on a single crystal
of the $S$=1 quasi-one-dimensional Heisenberg antiferromagnet NDMAP
and observed ESR spectra corresponding to the gapped state above the
critical field where the Haldane gap closes.  For $H \parallel a$ 
our ESR data above $H_{\rm c}$, taken at
$T=1.5$~K, agree very well with those from inelastic neutron
scattering \cite{zheludev} obtained at much lower temperature, and the
observed resonances are not affected by the boundary of the
long-range-ordered phase at $H\simeq9$~T.  For $H\parallel c$ 
we observed a change in slope of the ESR
line resulting from an avoided crossing between the two modes
having nontrivial slopes $H$ and $2H$.  
The observed field dependence of the magnon modes for both
geometries $H\parallel a$ and $H\parallel c$ is very different from
that expected from a conventional spin-wave theory.  Model
calculations based on the phenomenological field-theoretical approach
have reproduced all the features of the observation and a satisfactory
quantitative agreement between theory and experiment is
obtained.


\emph{Acknowledgments.---}
We 
would like to 
thank A. Zheludev for enlightening discussions.
This work was in part supported by the Molecular Ensemble research 
program from RIKEN and the Grant-in-Aid for Scientific 
Research on Priority Areas(B): Field-Induced New Quantum Phenomena 
in Magnetic Systems (No.13130203) from the Japanese
Ministry of Education, Culture, Sports, Science and Technology, and 
by the Grant I/75895 
``Low-dimensional magnets in high magnetic fields'' 
from Volkswagen-Stiftung.  


\end{document}